\documentclass[aps,pra,twocolumn,superscriptaddress,floatfix,longbibliography]{revtex4-2}

\usepackage{amsmath,amssymb,bm}
\usepackage{graphicx}
\usepackage{url}
\usepackage{longtable}
\usepackage{tikz}
\usepackage{pgfplots}
\pgfplotsset{compat=1.16}
\usetikzlibrary{positioning,arrows.meta}

\pgfplotsset{
  d2curves/.style={
    cycle list={%
      {black,          solid}, {black,          densely dashed},
      {red!75!black,   solid}, {red!75!black,   densely dashed},
      {blue!70!black,  solid}, {blue!70!black,  densely dashed},
      {teal!80!black,  solid}, {teal!80!black,  densely dashed}},
    every axis plot/.append style={line width=0.45pt, mark=none},
  }
}

\newcommand{\Fg}{F_g}
\newcommand{\Fe}{F_e}
\newcommand{\mg}{m_g}
\newcommand{\me}{m_e}
\newcommand{\ket}[1]{\left|#1\right\rangle}
\newcommand{\bra}[1]{\left\langle#1\right|}

\begin{document}

\title{Cancellation of $D_2$ line transitions of alkali-metal atoms\\ by magnetic-field values}

\author{Artur Aleksanyan}
\email{arthuraleksan@gmail.com}
\affiliation{Institute for Physical Research, NAS of Armenia, Ashtarak-2 0203, Armenia}

\author{Susanna Petrosyan}
\affiliation{Independent Researcher}

\author{Emil Gazazyan}
\affiliation{Institute for Physical Research, NAS of Armenia, Ashtarak-2 0203, Armenia}
\affiliation{Institute for Informatics and Automation Problems, NAS of Armenia, Yerevan 0014, Armenia}

\date{\today}

\begin{abstract}
In a previous work the $\pi$ transitions of the $D_1$ line of alkali-metal atoms were shown to
cancel at magnetic-field values given by a single closed-form expression. The $D_2$ line has
until now resisted the same treatment, because the fixed-$m$ Hamiltonian blocks of the $^2P_{3/2}$
manifold reach the dimension $4\times4$, and the corresponding formulas were considered to be too
heavy to be useful. In this work we show that this difficulty is only apparent. Since the Zeeman
interaction couples only levels with $\Delta F=\pm1$, every block, whatever its dimension, is
tridiagonal; its characteristic polynomial therefore obeys a three-term recursion, and the
components of its eigenvectors are polynomials in the eigenvalue. Using these two properties we
obtain the eigenvalues in closed form by Ferrari's and Cardano's formulas, the eigenvectors without
any further diagonalization, and finally a single relation which gives the magnetic-field value
canceling a $D_2$ transition as an explicit function of the excited-state eigenvalue, the nuclear
spin $I$, the ground-state hyperfine splitting and the magnetic quantum number. A necessary
condition on $m$ and on the polarization is derived, which replaces the selection rule known for
the $D_1$ line. All the magnetic-field values canceling $\pi$, $\sigma^+$ and $\sigma^-$
transitions of $^{23}$Na, $^{39}$K, $^{40}$K, $^{41}$K, $^{85}$Rb, $^{87}$Rb and $^{133}$Cs are
calculated up to 20~kG and given with their uncertainties; there are 234 of them. Contrary to the
$D_1$ line, the $\sigma^{\pm}$ transitions of the $D_2$ line do cancel, and they account for the
majority of the values found. The accuracy of the calculated $B$ values is limited only by the
uncertainties of the excited-state hyperfine constants.
\end{abstract}

\maketitle

\section{Introduction}

Alkali-metal vapors are widely used in atomic physics, e.g., in laser experiments \cite{Sargsyan2015},
information storage \cite{Legaie2018}, spectroscopy \cite{Aleksanyan2020a,Auzinsh2010,Sargsyan2019},
magnetometry \cite{Yin2021,Budker2007} and laser frequency stabilization \cite{Talker2020}, and are
also the main material used to study Bose--Einstein condensates \cite{Hans2021}. This is due to the
fact that alkali-metal atoms have a high transition intensity close to the infrared range, where
continuous wave narrowband diode lasers have good features and are cheap, which makes the experiments
easier. These properties make the study of alkali-metal vapor transitions in an external magnetic
field very important.

It is well known that in a moderate external magnetic field $B$ the atomic energy levels split into
magnetic sublevels (Zeeman splitting), and that the frequency differences between ground and excited
sublevels deviate greatly from the linear behavior \cite{Tremblay1990,Papageorgiou1994,Sargsyan2008}
before the linear regime is re-established in the hyperfine Paschen--Back domain
\cite{Sargsyan2014,Sargsyan2018}. Significant changes also occur for the atomic transition
probabilities. For small values of $B$ (up to approximately 1000~G) the Zeeman split hyperfine
transitions overlap because of the Doppler broadening, so that sub-Doppler techniques must be used to
follow each transition individually \cite{Khanbekyan2016,Sargsyan2019}. It was demonstrated
\cite{Sargsyan2017} that a strong line narrowing can be achieved by derivative selective reflection
from a nanocell.

Among the modifications caused by the field, the most spectacular one is the complete cancellation of
individual transitions: for particular values of $B$ the transition probability between two given
magnetic sublevels becomes exactly zero. Such values were first tabulated numerically for the $D_1$
and $D_2$ lines of rubidium \cite{Aleksanyan2020b,Momier2020}, and were then obtained analytically
for the $D_1$ line of all alkali-metal atoms \cite{Aleksanyan2022}, where a single formula gives $B$
as a function of the nuclear spin, the magnetic quantum number and the hyperfine splittings. These
$B$ values are attractive because they do not depend on the temperature, on the vapor density, on the
laser power or on any other experimental parameter: they are fixed by atomic constants alone, and can
therefore serve as standards for the calibration of magnetometers.

For the $D_2$ line the situation has been different. In Ref.~\cite{Aleksanyan2020b} it was stated
that ``for matrices with a size over $2\times2$, analytical formulas are heavy, and we have performed
numerical calculations,'' and in Ref.~\cite{Momier2020} that ``since cancelled transitions involve
$3\times3$ or $4\times4$ blocks, we do not derive any analytical formula although it should be
possible based on Ferrari and Cardano's formulas.'' A stronger statement is found in
Ref.~\cite{Ciampini2017}, where one reads that for the excited state ``no analytical formula exist
for the eigenenergies, to be derived by diagonalizing numerically the Hamiltonian.'' To the best of
our knowledge, no work has so far given the $D_2$ cancellation fields analytically.

In the present paper we show that the $D_2$ problem does close in a form which is compact enough to
be written down and used. The key observation, which does not seem to have been exploited before in
this context, is that the Zeeman Hamiltonian connects only hyperfine levels differing by
$\Delta F=\pm1$, so that in the coupled basis ordered by increasing $F$ every fixed-$m$ block is
\emph{tridiagonal}. Two consequences follow immediately. First, the characteristic polynomial of the
block is a continuant and obeys a three-term recursion, which produces the quartic and the cubic in a
form where Ferrari's and Cardano's formulas can be applied directly. Second, the components of the
eigenvectors are, up to a common factor, the leading principal minors of the same matrix, so that no
diagonalization is needed once the eigenvalue is known. Combining the two we obtain the modified
transfer coefficient as an explicit function of the excited eigenvalue, and the cancellation condition
reduces to one scalar equation which can be solved for $B$. We may add that the same $4\times4$
eigenvalue problem, that of a spin $3/2$ subject to a Zeeman and a quadrupole interaction, was solved
exactly long ago in the context of nuclear magnetic resonance \cite{Muha1983}. The algebra is
therefore not exotic; it simply does not seem to have been carried over to the optical $D_2$ problem.

The paper is organized as follows. In Sec.~\ref{sec:theory} we build the Hamiltonian, establish the
tridiagonal structure and give the eigenvalues and eigenvectors in closed form. In
Sec.~\ref{sec:cancel} we derive the relation giving the magnetic-field values which cancel the
transitions, discuss the degenerate case of the stretched ground sublevels, and obtain the condition
on $m$ and on the polarization. Section~\ref{sec:results} contains the verification of the model, a
detailed discussion of $^{87}$Rb, the complete set of $B$ values with their uncertainties, and a
comparison with what is known for the $D_1$ line. The full list of the 234 cancellation fields is
given in the Appendix.

\section{Theory}\label{sec:theory}

\subsection{Hamiltonian and block structure}

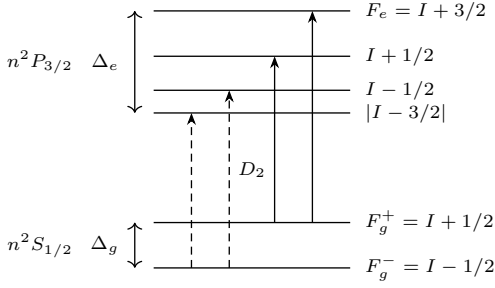
\begin{figure}[htbp]
\centering
\begin{tikzpicture}[x=1mm,y=1mm,line width=0.5pt,font=\scriptsize]
\foreach \y/\lab in {34/{$F_e=I+3/2$}, 28/{$I+1/2$}, 23.5/{$I-1/2$}, 20.5/{$|I-3/2|$}}
  { \draw (0,\y) -- (26,\y); \node[anchor=west] at (26.8,\y) {\lab}; }
\draw[<->] (-2.5,20.5) -- (-2.5,34);
\node[anchor=east] at (-3.4,27.25) {$\Delta_e$};
\node[anchor=east] at (-9.5,27.25) {$n^2P_{3/2}$};
\foreach \y/\lab in {6/{$F_g^{+}=I+1/2$}, 0/{$F_g^{-}=I-1/2$}}
  { \draw (0,\y) -- (26,\y); \node[anchor=west] at (26.8,\y) {\lab}; }
\draw[<->] (-2.5,0) -- (-2.5,6);
\node[anchor=east] at (-3.4,3) {$\Delta_g$};
\node[anchor=east] at (-9.5,3) {$n^2S_{1/2}$};
\draw[-{Stealth[length=1.6mm]},densely dashed] (5,0) -- (5,20.5);
\draw[-{Stealth[length=1.6mm]},densely dashed] (10,0) -- (10,23.5);
\draw[-{Stealth[length=1.6mm]}] (16,6) -- (16,28);
\draw[-{Stealth[length=1.6mm]}] (21,6) -- (21,34);
\node[fill=white,inner sep=1pt] at (13,13) {$D_2$};
\end{tikzpicture}
\caption{Scheme of the $D_2$ line when $I$ is a half-integer quantity. $\Delta_g$ is the
ground-state hyperfine splitting and $\Delta_e$ the total excited-state splitting. When $I$ is an
integer (here only $^{40}$K) the hyperfine structure is inverted in both manifolds, and the ordering
of the levels drawn in the figure is reversed.}
\label{fig:scheme}
\end{figure}

The fine structure results from the coupling between the orbital angular momentum $\mathbf{L}$ and
the spin angular momentum $\mathbf{S}$ of the single optical electron, the total electron angular
momentum being
\begin{equation}
\mathbf{J}=\mathbf{L}+\mathbf{S}.
\end{equation}
The hyperfine structure results from the coupling of $\mathbf{J}$ with the total nuclear angular
momentum $\mathbf{I}$,
\begin{equation}
\mathbf{F}=\mathbf{I}+\mathbf{J}.
\end{equation}
For the $D_2$ line one has $L=0$, $S=1/2$, $J_g=1/2$ for the ground state and $L=1$, $S=1/2$,
$J_e=3/2$ for the excited state. The total atomic angular momentum then takes the values
\begin{equation}
I-\tfrac12\leqslant \Fg\leqslant I+\tfrac12 ,\qquad
\left|I-\tfrac32\right|\leqslant \Fe\leqslant I+\tfrac32 ,
\end{equation}
and the magnetic quantum number obeys $-F\leqslant m\leqslant F$. This is the essential difference
with the $D_1$ line: the ground manifold still contains two hyperfine levels, but the excited
manifold contains four of them (three when $I=1/2$, which does not occur for the isotopes considered
here). Figure~\ref{fig:scheme} shows the level scheme.

In a static magnetic field the Hamiltonian is the sum of the unperturbed Hamiltonian and the Zeeman
Hamiltonian. We take the quantization axis along the field \cite{Tremblay1990}. In the unperturbed
basis $\ket{F,m}$ the diagonal elements are
\begin{equation}
\bra{F,m}\hat{H}\ket{F,m}=E_0(F)-\mu_B\,g_F(F)\,m\,B ,
\label{eq:diag}
\end{equation}
where $\mu_B$ is the Bohr magneton and $g_F(F)$ the associated Land\'e factor. For the $D_2$ line the
zero-field energies must include the electric quadrupole term, which is absent for $J=1/2$:
\begin{equation}
E_0(F)=\frac{A}{2}K+B_Q\,
\frac{\tfrac32K(K+1)-2I(I+1)J(J+1)}{2I(2I-1)\,2J(2J-1)} ,
\label{eq:E0}
\end{equation}
with $K=F(F+1)-I(I+1)-J(J+1)$, $A$ the magnetic dipole and $B_Q$ the electric quadrupole hyperfine
constant. The nondiagonal elements are
\begin{align}
\bra{F,m}\hat{H}\ket{F-1,m}&=\bra{F-1,m}\hat{H}\ket{F,m}\nonumber\\
&=-\frac{\mu_B}{2}\left(g_J-g_I\right)B\,\Omega(F,m),
\label{eq:offdiag}
\end{align}
where $g_J$ and $g_I$ are the total electronic and the nuclear Land\'e factors \cite{Bethe1957} and
\begin{align}
\Omega^2(F,m)={}&\frac{\left[(I+J+1)^2-F^2\right]\left[F^2-(I-J)^2\right]}{F^2}\nonumber\\
&\times\frac{F^2-m^2}{(2F+1)(2F-1)} .
\label{eq:omega}
\end{align}
For $J=1/2$ the factor $\Omega$ of Eq.~\eqref{eq:omega} reduces to $\sqrt{1-[2m/(1+2I)]^2}$, which is
the expression used for the $D_1$ line in Ref.~\cite{Aleksanyan2022}; for $J=3/2$ it does not
simplify, and it is precisely this factor which carries the whole $F$ dependence of the coupling.

The point which makes the problem tractable is that Eq.~\eqref{eq:offdiag} is the \emph{only}
nondiagonal element: the Zeeman operator is a rank-one tensor in $\mathbf{F}$ space and connects
$\ket{F,m}$ to $\ket{F\pm1,m}$ only. Consequently, if for a given $m$ the levels $F$ that contain
this $m$ are ordered by increasing $F$ and labelled $k=1,\dots,n$, the Hamiltonian block is the real
symmetric tridiagonal matrix
\begin{equation}
H^{(m)}=
\begin{pmatrix}
D_1 & W_1 & & \\
W_1 & D_2 & \ddots & \\
& \ddots & \ddots & W_{n-1}\\
& & W_{n-1} & D_n
\end{pmatrix},
\label{eq:block}
\end{equation}
where $D_k$ is given by Eq.~\eqref{eq:diag} and $W_k$ by Eq.~\eqref{eq:offdiag} evaluated at the
larger of the two coupled $F$ values. Since a level $F$ contains the sublevel $m$ only if
$F\geqslant|m|$, the dimension of the block is
\begin{equation}
n_g=\begin{cases}2, & |\mg|\leqslant I-\tfrac12\\[2pt] 1, & |\mg|=I+\tfrac12\end{cases}
\qquad
n_e=\begin{cases}4, & |\me|\leqslant I-\tfrac32\\ 3, & |\me|=I-\tfrac12\\ 2, & |\me|=I+\tfrac12\\
1, & |\me|=I+\tfrac32 .\end{cases}
\label{eq:dims}
\end{equation}
The ground-state blocks are the same $2\times2$ matrices as for the $D_1$ line and are diagonalized
by the Breit--Rabi formula. The whole difficulty of the $D_2$ line is contained in the excited blocks
of dimension 3 and 4.

\subsection{Eigenvalues}

For a tridiagonal matrix the characteristic polynomial is a continuant. Writing
$\chi_k(\lambda)=\det\!\left[\lambda\,\mathbf{1}-H^{(m)}\right]_{k\times k}$ for the leading principal
minor of order $k$, one has $\chi_0=1$, $\chi_1=\lambda-D_1$ and
\begin{equation}
\chi_k(\lambda)=\left(\lambda-D_k\right)\chi_{k-1}(\lambda)-W_{k-1}^2\,\chi_{k-2}(\lambda),
\qquad k\geqslant2 .
\label{eq:continuant}
\end{equation}
The eigenvalues are the roots of $\chi_n(\lambda)=0$. Only the squares $W_k^2$ enter, so that the
sign convention chosen in Eq.~\eqref{eq:offdiag} is immaterial for the spectrum.

For $n=4$, Eq.~\eqref{eq:continuant} applied to the block \eqref{eq:block} gives the quartic
$\lambda^4+c_3\lambda^3+c_2\lambda^2+c_1\lambda+c_0=0$ with
\begin{align}
c_3&=-s_1,\nonumber\\
c_2&=s_2-\left(W_1^2+W_2^2+W_3^2\right),\nonumber\\
c_1&=-s_3+W_1^2(D_3+D_4)+W_2^2(D_1+D_4)\nonumber\\
&\quad+W_3^2(D_1+D_2),\nonumber\\
c_0&=s_4-W_1^2D_3D_4-W_2^2D_1D_4-W_3^2D_1D_2\nonumber\\
&\quad+W_1^2W_3^2 ,
\label{eq:quartic}
\end{align}
where $s_1,\dots,s_4$ are the elementary symmetric polynomials of $D_1,\dots,D_4$. The depressed form
is obtained with $\lambda=y-c_3/4$, giving $y^4+py^2+qy+r=0$ with
\begin{align}
p&=c_2-\tfrac38c_3^2,\nonumber\\
q&=c_1-\tfrac12c_2c_3+\tfrac18c_3^3,\nonumber\\
r&=c_0-\tfrac14c_1c_3+\tfrac1{16}c_2c_3^2-\tfrac3{256}c_3^4 .
\end{align}
where the coefficients \eqref{eq:quartic} are written directly in terms of the matrix entries of
Eq.~\eqref{eq:block}, the diagonal ones being given by Eqs.~\eqref{eq:diag} and \eqref{eq:E0}.
Ferrari's method then requires one real root $z$ of the resolvent cubic
\begin{equation}
z^3-\tfrac{p}{2}z^2-r\,z+\left(\tfrac{rp}{2}-\tfrac{q^2}{8}\right)=0 ,
\label{eq:resolvent}
\end{equation}
which is obtained by Cardano's formula; with any real root of Eq.~\eqref{eq:resolvent} the four
eigenvalues follow from
\begin{equation}
y=\frac{\pm_1\sqrt{2z-p}\;\pm_2\sqrt{-\!\left(2z+p\pm_1\dfrac{2q}{\sqrt{2z-p}}\right)}}{2},
\label{eq:ferrari}
\end{equation}
the two signs being chosen independently. For $n=3$ the characteristic polynomial is the cubic
\begin{equation}
\lambda^3-s_1\lambda^2+\left[s_2-W_1^2-W_2^2\right]\lambda
-\left[s_3-W_1^2D_3-W_2^2D_1\right]=0 ,
\label{eq:cubic}
\end{equation}
solved by Cardano's formula, and for $n=2$ one recovers the Breit--Rabi surd
\begin{equation}
\lambda_\pm=\frac{D_1+D_2}{2}\pm\sqrt{\left(\frac{D_1-D_2}{2}\right)^2+W_1^2}.
\label{eq:br}
\end{equation}
Since $H^{(m)}$ is real symmetric and tridiagonal with $W_k\neq0$ for $B\neq0$, its eigenvalues are
real and simple, and they never cross as functions of $B$. This last property is what allows the
eigenstates to be labelled unambiguously: each eigenvalue is identified by the rank of the zero-field
hyperfine energy it originates from. It should be noted that labelling instead by the order of $F$
would be wrong for $^{40}$K, whose hyperfine structure is inverted in both manifolds.

\subsection{Eigenvectors}

The tridiagonal structure gives the eigenvectors as well, without any further work. Let
$\lambda$ be an eigenvalue of $H^{(m)}$ and let $\mathbf{c}=(c_1,\dots,c_n)^{\mathsf T}$ be the
associated eigenvector expressed in the unperturbed basis $\ket{F_k,m}$. Writing the eigenvalue
equations row by row and eliminating recursively one finds
\begin{equation}
c_k(\lambda)=\mathcal{N}\,\chi_{k-1}(\lambda)\prod_{j=k}^{n-1}W_j ,
\qquad k=1,\dots,n ,
\label{eq:evec}
\end{equation}
where $\mathcal{N}$ is fixed by normalization and an empty product is equal to one. In other words,
once the eigenvalue is known the eigenvector components are the continuants \eqref{eq:continuant}
themselves, evaluated at that eigenvalue. Equation~\eqref{eq:evec} holds for any $n$ and replaces the
$2\times2$ closed forms used for the $D_1$ line. For the ground state, $n_g=2$ and Eq.~\eqref{eq:evec}
reduces to $\mathbf{c}^{\,g}\propto(W_1,\,\lambda_g-D_1)$.

\subsection{Transfer coefficients}

The intensity of the transition between two magnetic sublevels is proportional to the square of the
transfer coefficient. For the unperturbed states the coefficient reads \cite{Auzinsh2010}
\begin{align}
a\!\left(\Fe,\me;\Fg,\mg;q\right)
={}&(-1)^{1+I+J_e+\Fe+\Fg-\me}\nonumber\\
&\times\sqrt{(2J_e+1)(2\Fe+1)(2\Fg+1)}\nonumber\\
&\times
\begin{pmatrix}\Fe&1&\Fg\\-\me&q&\mg\end{pmatrix}
\begin{Bmatrix}\Fe&1&\Fg\\ J_g&I&J_e\end{Bmatrix},
\label{eq:aun}
\end{align}
where $q=0,\pm1$ corresponds to $\pi$, $\sigma^{\pm}$ excitation and $J_e=3/2$. In the presence of the
field the atomic states are the eigenvectors of the blocks, and the transfer coefficient becomes the
double contraction
\begin{equation}
a^{\rm mod}=\sum_{j=1}^{n_e}\sum_{k=1}^{n_g}
c^{\,e}_j(\lambda_e)\;a\!\left(F^{e}_j,\me;F^{g}_k,\mg;q\right)\,c^{\,g}_k(\lambda_g),
\label{eq:amod}
\end{equation}
which we call, following Ref.~\cite{Aleksanyan2022}, the modified transfer coefficient. Both
$c^{\,e}$ and $c^{\,g}$ are given by Eq.~\eqref{eq:evec}, so that $a^{\rm mod}$ is an explicit
algebraic function of $B$.

\section{Magnetic-field values canceling the transitions}\label{sec:cancel}

\subsection{The general relation}

A transition is canceled when $a^{\rm mod}=0$. The double sum \eqref{eq:amod} contains the two
manifolds in a very asymmetric way, and it is this asymmetry which produces a closed form. Let us
first perform the sum over the excited index only, and define the two contraction polynomials
\begin{equation}
U(\lambda_e)=\sum_{j=1}^{n_e}c^{\,e}_j\,a\!\left(F^{e}_j;\Fg^{-}\right),\quad
V(\lambda_e)=\sum_{j=1}^{n_e}c^{\,e}_j\,a\!\left(F^{e}_j;\Fg^{+}\right),
\label{eq:UV}
\end{equation}
where $\Fg^{\mp}=I\mp1/2$ are the two ground hyperfine levels and the remaining arguments of $a$ are
understood. Equation~\eqref{eq:amod} then reads simply
\begin{equation}
a^{\rm mod}=c^{\,g}_1U(\lambda_e)+c^{\,g}_2V(\lambda_e),
\end{equation}
so that the cancellation condition is a condition on the \emph{ground-state} eigenvector alone,
\begin{equation}
\frac{c^{\,g}_1}{c^{\,g}_2}=-\frac{1}{\rho},
\qquad
\rho\equiv\frac{U(\lambda_e)}{V(\lambda_e)} .
\label{eq:rho}
\end{equation}
The ratio $\rho$ contains all the information about the $4\times4$ or $3\times3$ excited block, but it
enters the problem as one scalar only.

The ground block is a $2\times2$ matrix, and its eigenvector ratio $r=c^{\,g}_1/c^{\,g}_2$ satisfies
the elementary relation
\begin{equation}
D_2-D_1=W_1\,\frac{1-r^2}{r} .
\label{eq:ratio}
\end{equation}
For the ground state one has, setting $E_0(\Fg^{-})=0$,
\begin{align}
D_2-D_1&=\Delta_g-\frac{2\mu_B\,\sigma\,\mg B}{2I+1},&\quad \Delta_g&=A_g\left(I+\tfrac12\right),\nonumber\\
W_1&=-\frac{\mu_B}{2}\,\sigma B\beta, &\quad \sigma&=g_S-g_I,
\label{eq:gs}
\end{align}
with $\beta=\sqrt{1-\left[2\mg/(2I+1)\right]^2}$. Substituting $r=-1/\rho$ from Eq.~\eqref{eq:rho}
into Eq.~\eqref{eq:ratio} and using Eq.~\eqref{eq:gs}, the field factorizes out and one obtains the
central result of this work,
\begin{equation}
\boxed{\;
B=\frac{\rho\,\Delta_g}
{-\mu_B\left(g_S-g_I\right)
\left[\dfrac{\beta}{2}\left(1-\rho^2\right)-\dfrac{2\mg\,\rho}{2I+1}\right]}\;}
\label{eq:master}
\end{equation}
Equation~\eqref{eq:master} is the $D_2$ counterpart of Eq.~(17) of Ref.~\cite{Aleksanyan2022}. It
expresses the canceling field as a function of the nuclear spin, of the ground-state hyperfine
splitting, of the Land\'e factors and of the magnetic quantum number, exactly as in the $D_1$ case;
the whole effect of the larger excited manifold is carried by the single number $\rho$, which is
obtained from Eqs.~\eqref{eq:evec}, \eqref{eq:aun} and \eqref{eq:UV} once $\lambda_e$ is known from
Eq.~\eqref{eq:ferrari} or \eqref{eq:cubic}. Since $\lambda_e$ itself depends on $B$,
Eq.~\eqref{eq:master} is an implicit relation; in practice it is used either as a fixed point or,
equivalently, by eliminating $\lambda$ between $\chi_{n_e}(\lambda)=0$ and Eq.~\eqref{eq:master},
which yields a single polynomial equation in $B$ whose roots are the canceling fields.

We note in passing that when $\mg=0$ the second term in the bracket disappears and
Eq.~\eqref{eq:master} takes the simple form $B=2\rho\Delta_g/\left[-\mu_B\sigma(1-\rho^2)\right]$,
since $\beta=1$ in that case.

\subsection{Degenerate case}

Equation~\eqref{eq:master} assumes $n_g=2$. When $|\mg|=I+1/2$ the ground block is one-dimensional,
the ground state is the pure state $\ket{I+1/2,\mg}$, $\beta=0$ and $W_1=0$; the derivation above
breaks down because $U\equiv0$. In that case the modified transfer coefficient reduces to
$a^{\rm mod}=V(\lambda_e)$ and the cancellation condition becomes simply
\begin{equation}
V\!\left(\lambda_e(B)\right)=0 .
\label{eq:degen}
\end{equation}
This is a genuine branch and not an artefact: six of the cancellation fields reported below belong to
it. All of them have $n_g=1$ and $n_e=3$, since $|\me|=|\mg\pm1|=I-1/2$ in this configuration.

\subsection{Condition on $m$ and on the polarization}

Not every combination of $m$ and $q$ can produce a cancellation, and the admissible ones are fixed by
Eq.~\eqref{eq:dims}. Two cases must be excluded.

For $\sigma^{q}$ excitation with $\mg=q\left(I+1/2\right)$ one has $\me=q\left(I+3/2\right)$, so that
both blocks are one-dimensional. The transition is the cycling transition, $a^{\rm mod}$ is equal to
its unperturbed value for every $B$, and no cancellation can occur.

For $\pi$ excitation with $|m|=I+1/2$ the ground block is one-dimensional and the excited block is
$2\times2$, spanned by $F=I+1/2$ and $F=I+3/2$. The cancellation would require the excited eigenvector
ratio $t=c^{\,e}_1/c^{\,e}_2$ to take the value
\begin{equation}
t^{\star}=-\frac{a\!\left(I+\tfrac32;I+\tfrac12\right)}{a\!\left(I+\tfrac12;I+\tfrac12\right)}
=\mathrm{sgn}(m)\sqrt{\frac{3}{2I}} ,
\label{eq:tstar}
\end{equation}
the last equality in Eq.~\eqref{eq:tstar} being an exact identity of the $3j$ and $6j$ symbols
entering Eq.~\eqref{eq:aun} at $\me=\mg=m=\pm(I+1/2)$ and $q=0$. Of the two eigenvectors of the
$2\times2$ excited block, one has a
ratio which remains at a finite distance from $t^{\star}$, and the other tends to $t^{\star}$ only in
the limit $B\to\infty$, the difference decreasing as $O(1/B)$ without ever changing sign. We have
verified this behavior for all seven isotopes and both signs of $m$ over the range
$10^{-2}$--$10^{9}$~G. Consequently the transition becomes arbitrarily weak at high field but is
never canceled at any finite value of $B$.

Collecting the two cases, a necessary condition for a $D_2$ cancellation is
\begin{equation}
\left|m\right|\leqslant I-\tfrac12\quad(\pi),
\qquad
\mg\neq q\left(I+\tfrac12\right)\quad(\sigma^{q}).
\label{eq:selrule}
\end{equation}
Condition \eqref{eq:selrule} should be compared with Eq.~(18) of Ref.~\cite{Aleksanyan2022}, which for
the $D_1$ line reads $0\leqslant(-1)^{2I}m\leqslant I-1/2$ and selects one sign of $m$ only. On the
$D_2$ line the parity restriction disappears: both signs of $m$ produce cancellations, at different
fields. The condition is necessary but not sufficient. Two configurations satisfying
Eq.~\eqref{eq:selrule} have no root at all, namely $^{40}$K $\sigma^{+}$ at $\mg=-9/2$ and $^{133}$Cs
$\sigma^{+}$ at $\mg=-4$, for which we found no solution up to 400~kG.

\section{Results and discussion}\label{sec:results}

\subsection{Physical quantities and verification}

The isotope data used in the calculations are collected in Table~\ref{tab:const}. We considered
$^{23}$Na, $^{39}$K, $^{40}$K, $^{41}$K, $^{85}$Rb, $^{87}$Rb and $^{133}$Cs; all of them are stable
except $^{40}$K and $^{87}$Rb, whose half-lives are $1.248(3)\times10^{9}$ and
$49.23(22)\times10^{9}$ years, respectively. As in Ref.~\cite{Aleksanyan2022} we used
$\mu_B/h=-1.399\,624\,504\,2(86)$~MHz/G and $g_S=2.002\,319\,304\,362\,2(15)$ \cite{Mohr2016}, and
$g_L$ was calculated with the exact formula of Phillips \cite{Phillips1949} using the isotope masses
of Ref.~\cite{Audi2003}. For $J_e=3/2$ the electronic Land\'e factor is
$g_J=\left(2g_L+g_S\right)/3=1.334\,102$.

\begin{table*}[t]
\caption{Values used to calculate the transitions between $D_2$ line magnetic sublevels, with their
uncertainties. $A_g$ is the ground-state magnetic dipole constant, $A_e$ and $B_Q$ the excited-state
magnetic dipole and electric quadrupole constants. The value of $A_g$ for $^{133}$Cs is exact by the
definition of the second. Data are taken from Ref.~\cite{Steck} for Na, Rb and Cs and from
Ref.~\cite{Tiecke} for the three potassium isotopes.}
\label{tab:const}
\begin{ruledtabular}
\begin{tabular}{lccccc}
Isotope & $I$ & $g_I$ & $A_g$ (MHz) & $A_e$ (MHz) & $B_Q$ (MHz)\\
\hline
$^{23}$Na  & 3/2 & $-0.000\,804\,611$ & 885.813\,064\,40   & 18.534(15)   & 2.724(30)\\
$^{39}$K   & 3/2 & $-0.000\,141\,935$ & 230.859\,860\,1    & 6.093(25)    & 2.786(71)\\
$^{40}$K   & 4   & $+0.000\,176\,490$ & $-285.7308(24)$    & $-7.585(10)$ & $-3.445(90)$\\
$^{41}$K   & 3/2 & $-0.000\,077\,906$ & 127.006\,935\,2    & 3.363(14)    & 3.351(96)\\
$^{85}$Rb  & 5/2 & $-0.000\,293\,640$ & 1011.910\,813\,0   & 25.0020(99)  & 25.790(93)\\
$^{87}$Rb  & 3/2 & $-0.000\,995\,141$ & 3417.341\,305\,45  & 84.7185(20)  & 12.4965(37)\\
$^{133}$Cs & 7/2 & $-0.000\,398\,854$ & 2298.157\,942\,5   & 50.288\,27(23) & $-0.4934(17)$\\
\end{tabular}
\end{ruledtabular}
\end{table*}

The model was verified in three independent ways. First, the eigenvalues and eigenvectors obtained
from Eqs.~\eqref{eq:ferrari}, \eqref{eq:cubic}, \eqref{eq:br} and \eqref{eq:evec} were compared with
a numerical diagonalization of the same blocks; over 1150 eigenvalue and 3550 eigenvector
evaluations the largest relative deviations were $4.2\times10^{-14}$ and $4.4\times10^{-16}$,
i.e.\ at the level of the machine precision. Second, the fields predicted by Eq.~\eqref{eq:master}
were compared with the roots obtained by numerically minimizing $\left(a^{\rm mod}\right)^2$: over the
228 fields belonging to the generic branch the worst relative deviation is $1.6\times10^{-11}$, and
over the six fields of the degenerate branch \eqref{eq:degen} it is $2.1\times10^{-13}$. Third, the
values obtained here were compared with the 58 $D_2$ fields of $^{85}$Rb and $^{87}$Rb published in
Ref.~\cite{Aleksanyan2020b}, which were computed there by a completely independent numerical route;
all 58 agree, the largest discrepancy being $0.84\sigma$. This last comparison is worth stressing,
because it shows that the closed form reproduces values which were previously accessible only
numerically.

\subsection{The $^{87}$Rb $D_2$ line}

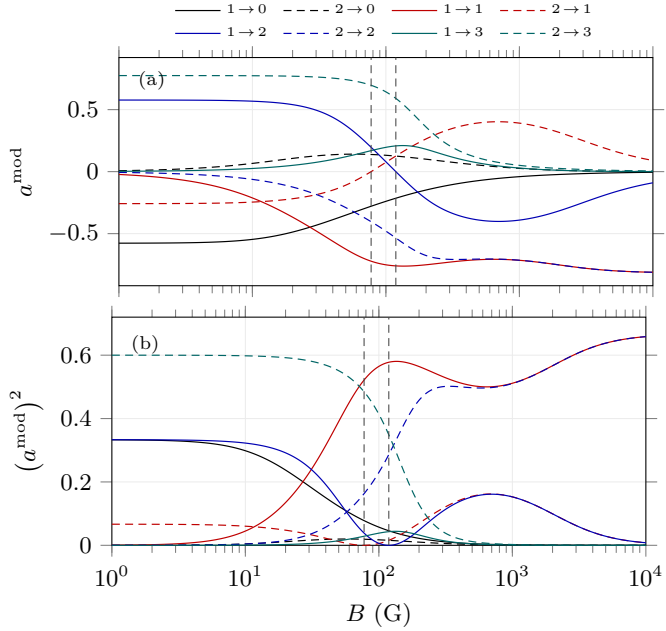
\begin{figure}[htbp]
\centering
\begin{tikzpicture}
\begin{semilogxaxis}[
  width=\columnwidth, height=4.6cm,
  xlabel={}, ylabel={$a^{\rm mod}$},
  xmin=1, xmax=10000, ymin=-0.92, ymax=0.92,
  xticklabels={}, tick align=outside, grid=major, grid style={gray!16, line width=0.3pt},
  axis line style={line width=0.4pt}, tick style={line width=0.4pt},
  legend style={font=\tiny, at={(0.5,1.03)}, anchor=south, legend columns=4,
                draw=none, fill=none, /tikz/every even column/.append style={column sep=4pt}},
  d2curves,
]
\addplot table[x index=0,y index=1] {fig_87Rb_pi_m0.dat}; \addlegendentry{$1\!\rightarrow\!0$}
\addplot table[x index=0,y index=2] {fig_87Rb_pi_m0.dat}; \addlegendentry{$2\!\rightarrow\!0$}
\addplot table[x index=0,y index=3] {fig_87Rb_pi_m0.dat}; \addlegendentry{$1\!\rightarrow\!1$}
\addplot table[x index=0,y index=4] {fig_87Rb_pi_m0.dat}; \addlegendentry{$2\!\rightarrow\!1$}
\addplot table[x index=0,y index=5] {fig_87Rb_pi_m0.dat}; \addlegendentry{$1\!\rightarrow\!2$}
\addplot table[x index=0,y index=6] {fig_87Rb_pi_m0.dat}; \addlegendentry{$2\!\rightarrow\!2$}
\addplot table[x index=0,y index=7] {fig_87Rb_pi_m0.dat}; \addlegendentry{$1\!\rightarrow\!3$}
\addplot table[x index=0,y index=8] {fig_87Rb_pi_m0.dat}; \addlegendentry{$2\!\rightarrow\!3$}
\draw[gray!55!black,densely dashed,line width=0.4pt] (axis cs:77.504,-0.92) -- (axis cs:77.504,0.92);
\draw[gray!55!black,densely dashed,line width=0.4pt] (axis cs:118.705,-0.92) -- (axis cs:118.705,0.92);
\node[font=\scriptsize,anchor=north west] at (axis cs:1.2,0.88) {(a)};
\end{semilogxaxis}
\end{tikzpicture}

\vspace{-1.5mm}

\begin{tikzpicture}
\begin{semilogxaxis}[
  width=\columnwidth, height=4.6cm,
  xlabel={$B$ (G)}, ylabel={$\left(a^{\rm mod}\right)^2$},
  xmin=1, xmax=10000, ymin=0, ymax=0.72,
  tick align=outside, grid=major, grid style={gray!16, line width=0.3pt},
  axis line style={line width=0.4pt}, tick style={line width=0.4pt},
  d2curves,
]
\addplot table[x index=0,y expr={\thisrowno{1}*\thisrowno{1}}] {fig_87Rb_pi_m0.dat};
\addplot table[x index=0,y expr={\thisrowno{2}*\thisrowno{2}}] {fig_87Rb_pi_m0.dat};
\addplot table[x index=0,y expr={\thisrowno{3}*\thisrowno{3}}] {fig_87Rb_pi_m0.dat};
\addplot table[x index=0,y expr={\thisrowno{4}*\thisrowno{4}}] {fig_87Rb_pi_m0.dat};
\addplot table[x index=0,y expr={\thisrowno{5}*\thisrowno{5}}] {fig_87Rb_pi_m0.dat};
\addplot table[x index=0,y expr={\thisrowno{6}*\thisrowno{6}}] {fig_87Rb_pi_m0.dat};
\addplot table[x index=0,y expr={\thisrowno{7}*\thisrowno{7}}] {fig_87Rb_pi_m0.dat};
\addplot table[x index=0,y expr={\thisrowno{8}*\thisrowno{8}}] {fig_87Rb_pi_m0.dat};
\draw[gray!55!black,densely dashed,line width=0.4pt] (axis cs:77.504,0) -- (axis cs:77.504,0.72);
\draw[gray!55!black,densely dashed,line width=0.4pt] (axis cs:118.705,0) -- (axis cs:118.705,0.72);
\node[font=\scriptsize,anchor=north west] at (axis cs:1.2,0.69) {(b)};
\end{semilogxaxis}
\end{tikzpicture}
\caption{(a) Modified transfer coefficients of the $^{87}$Rb $D_2$ line $\pi$ transitions with
$m=0$, as functions of the magnetic field, on a logarithmic scale from 1~G to 10~kG. The colour
indicates the excited level $\Fe$ and the line style the ground level $\Fg$ (solid, $\Fg=1$; dashed,
$\Fg=2$). (b) The same quantities squared. The vertical dashed lines mark the two cancellations of
this group, at $B=77.5040(24)$~G for $2\rightarrow1$ and $B=118.7049(33)$~G for $1\rightarrow2$.
Only these two coefficients change sign; the six others keep it over the whole range. Contrary to
the $D_1$ line, no other coefficient of the group is stationary at these fields. Above a few
kilogauss the curves flatten onto their hyperfine Paschen--Back values, which is why the scan for
cancellations can be stopped at 20~kG.}
\label{fig:87Rb}
\end{figure}

In order that the mechanism be well understood by the reader, we discuss $^{87}$Rb in some detail. For
$I=3/2$ the ground state has $\Fg=1,2$ and the excited state $\Fe=0,1,2,3$. Consider the $\pi$
transitions with $m=0$: the ground block is $2\times2$ and the excited block is $4\times4$, so eight
transitions are a priori possible. Figure~\ref{fig:87Rb} shows the corresponding modified transfer
coefficients as functions of $B$.

Two of the eight coefficients change sign, at $B=77.5040(24)$~G and $B=118.7049(33)$~G, and it is
only these two which are canceled. Let us follow the second one in detail, since it illustrates the
use of Eq.~\eqref{eq:master}. At $B=118.7049$~G the $4\times4$ excited block has the eigenvalue
$\lambda_e=+64.882$~MHz, which is the third one when the levels are ordered by their zero-field
energies; inserting it in Eq.~\eqref{eq:evec} and then in Eq.~\eqref{eq:UV} gives
$\rho=0.024\,334\,573$. With $\Delta_g=A_g(I+1/2)=6834.682\,61$~MHz, $\beta=1$ and $\mg=0$,
Eq.~\eqref{eq:master} returns $B=118.7049$~G, which is the value at which the coefficient was found
to vanish. The whole $4\times4$ problem has thus been compressed into the single number $\rho$.

The remaining transitions of $^{87}$Rb are collected in Table~\ref{tab:87Rb}. Comparison with Tables~5
and 6 of Ref.~\cite{Aleksanyan2020b} shows agreement everywhere within the combined uncertainties. It
should be noted that the entry at $B=211.1200(62)$~G is one of a small family discussed in
Sec.~\ref{sec:small} below.

\begin{table}[t]
\caption{Magnetic-field values canceling the $D_2$ line transitions of $^{87}$Rb, with their
uncertainties. $S$ is the peak intensity of the canceled transition in the neighborhood of the root,
relative to the strongest transition of the same group; transitions with $S\lesssim10^{-9}$ are not
observable in practice.}
\label{tab:87Rb}
\begin{ruledtabular}
\begin{tabular}{lcccrl}
 & Pol. & $\Fg\rightarrow\Fe$ & $\mg$ & $B$ (G)\hspace{4mm} & $S$\\
\hline
$^{87}$Rb & $\pi$ & $2\rightarrow2$ & $-1$ & 55.6960(13) & $1.6\times10^{-2}$\\
$^{87}$Rb & $\pi$ & $2\rightarrow1$ & 0 & 77.5040(24) & $1.2\times10^{-2}$\\
$^{87}$Rb & $\pi$ & $1\rightarrow2$ & 0 & 118.7049(33) & $1.3\times10^{-2}$\\
$^{87}$Rb & $\pi$ & $2\rightarrow1$ & 1 & 77.2390(24) & $3.8\times10^{-3}$\\
$^{87}$Rb & $\pi$ & $1\rightarrow2$ & 1 & 114.2387(33) & $3.8\times10^{-3}$\\
\hline
$^{87}$Rb & $\sigma^{+}$ & $2\rightarrow1$ & $-1$ & 37.7176(15) & $2.8\times10^{-2}$\\
$^{87}$Rb & $\sigma^{+}$ & $2\rightarrow1$ & 0 & 35.0307(14) & $8.7\times10^{-3}$\\
$^{87}$Rb & $\sigma^{+}$ & $2\rightarrow2$ & $-1$ & 157.6237(39) & $1.1\times10^{-2}$\\
$^{87}$Rb & $\sigma^{+}$ & $2\rightarrow2$ & 0 & 183.1465(47) & $2.3\times10^{-3}$\\
$^{87}$Rb & $\sigma^{+}$ & $2\rightarrow2$ & 1 & 211.1200(62) & $4.5\times10^{-4}$\\
$^{87}$Rb & $\sigma^{+}$ & $2\rightarrow1$ & $-2$ & 1792.12(55) & $2.5\times10^{-9}$\\
$^{87}$Rb & $\sigma^{+}$ & $1\rightarrow0$ & $-1$ & 1596.51(57) & $8.6\times10^{-10}$\\
$^{87}$Rb & $\sigma^{+}$ & $2\rightarrow0$ & $-1$ & 1761.14(58) & $4.1\times10^{-10}$\\
\hline
$^{87}$Rb & $\sigma^{-}$ & $1\rightarrow2$ & 1 & 71.9227(19) & $2.1\times10^{-3}$\\
$^{87}$Rb & $\sigma^{-}$ & $1\rightarrow1$ & 1 & 140.8254(58) & $9.0\times10^{-4}$\\
$^{87}$Rb & $\sigma^{-}$ & $1\rightarrow2$ & 0 & 114.3041(33) & $3.8\times10^{-3}$\\
\end{tabular}
\end{ruledtabular}
\end{table}

\subsection{All isotopes}

We have examined all the $\pi$, $\sigma^{+}$ and $\sigma^{-}$ transitions of the seven isotopes in
the range $0$--$20$~kG. In total 234 cancellation fields were found, distributed as shown in
Table~\ref{tab:census}. The complete list is given in the Appendix. Figure~\ref{fig:overview} shows
their distribution.

\begin{table}[htbp]
\caption{Number of $D_2$ cancellation fields below 20~kG for each isotope and polarization, and the
lowest and highest field found. The count follows the nuclear spin, as Eq.~\eqref{eq:dims} requires:
a larger $I$ means more values of $m$ and therefore more blocks large enough for the levels to
interfere.}
\label{tab:census}
\begin{ruledtabular}
\begin{tabular}{lcccccrr}
Isotope & $I$ & $\pi$ & $\sigma^{+}$ & $\sigma^{-}$ & Total & $B_{\min}$ (G) & $B_{\max}$ (G)\\
\hline
$^{23}$Na  & 3/2 & 7  & 8  & 5  & 20 & 7.62 & 12\,172\\
$^{39}$K   & 3/2 & 7  & 8  & 5  & 20 & 2.04 & 952\\
$^{40}$K   & 4   & 28 & 20 & 28 & 76 & 1.44 & 12\,192\\
$^{41}$K   & 3/2 & 7  & 7  & 5  & 19 & 0.64 & 212\\
$^{85}$Rb  & 5/2 & 15 & 16 & 11 & 42 & 6.57 & 6\,681\\
$^{87}$Rb  & 3/2 & 5  & 8  & 3  & 16 & 35.03 & 1\,792\\
$^{133}$Cs & 7/2 & 17 & 13 & 11 & 41 & 20.36 & 237\\
\hline
Total      &     & 86 & 80 & 68 & 234 & & \\
\end{tabular}
\end{ruledtabular}
\end{table}

\begin{figure}[htbp]
\centering
\begin{tikzpicture}
\begin{semilogxaxis}[
  width=\columnwidth, height=5.4cm,
  xlabel={$B$ (G)}, xmin=0.4, xmax=30000,
  ytick={1,2,3,4,5,6,7},
  yticklabels={$^{23}$Na,$^{39}$K,$^{40}$K,$^{41}$K,$^{85}$Rb,$^{87}$Rb,$^{133}$Cs},
  ymin=0.4, ymax=7.6, y dir=reverse,
  tick align=outside, grid=major, grid style={gray!18},
  legend style={font=\tiny, at={(0.5,1.02)}, anchor=south, draw=none, fill=none,
                legend columns=3, /tikz/every even column/.append style={column sep=5pt}},
]
\addplot[only marks,mark=*,mark size=1.1pt,black]
  table[x index=1,y expr={\thisrowno{0}-0.22}] {fig_overview_pi.dat};
\addlegendentry{$\pi$}
\addplot[only marks,mark=triangle*,mark size=1.4pt,red!75!black]
  table[x index=1,y index=0] {fig_overview_sp.dat};
\addlegendentry{$\sigma^{+}$}
\addplot[only marks,mark=square*,mark size=1.0pt,blue!70!black]
  table[x index=1,y expr={\thisrowno{0}+0.22}] {fig_overview_sm.dat};
\addlegendentry{$\sigma^{-}$}
\draw[gray,densely dashed] (axis cs:237,0.4) -- (axis cs:237,7.6);
\end{semilogxaxis}
\end{tikzpicture}
\caption{The 234 magnetic-field values canceling $D_2$ transitions below 20~kG, on a logarithmic
scale. Within each isotope the three polarizations are drawn on slightly displaced rows. The dashed
line at 237~G is the upper bound of the observable subset: every one of the 174 fields with
$S\geqslant10^{-4}$ lies below it, while all the values above it belong to transitions whose
intensity in the vicinity of the root is smaller than $10^{-4}$ of the strongest transition of their
group.}
\label{fig:overview}
\end{figure}
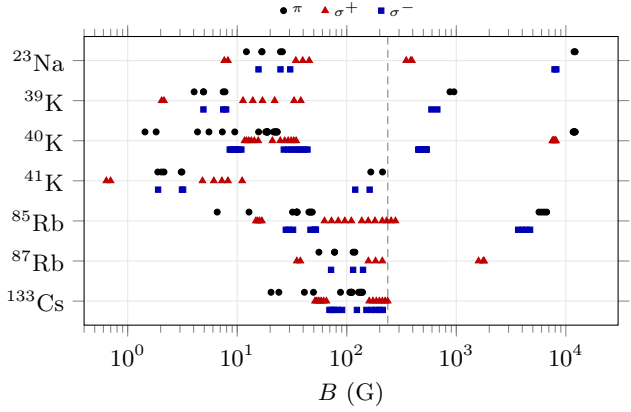

Three features deserve comment. First, the number of fields grows with $I$, as it must, since a larger
nuclear spin allows more values of $m$ to satisfy Eq.~\eqref{eq:selrule}; $^{40}$K alone carries a
third of the whole set. Second, the $\sigma^{\pm}$ transitions do cancel, and they are the majority:
148 of the 234 fields, against 86 for $\pi$. This is the sharpest qualitative difference with the
$D_1$ line, for which it was shown \cite{Aleksanyan2022,Aleksanyan2020b} that no $\sigma^{\pm}$
transition is ever canceled. Any treatment of the $D_2$ line that carries the $D_1$ result forward
would therefore miss most of the effect. Third, $\Delta F=\pm2$ transitions, which do not exist on the
$D_1$ line, contribute 35 fields; being forbidden at zero field, they borrow their whole strength from
the hyperfine mixing that the field itself destroys, and 27 of the 35 are consequently far too weak to
be measured.

This last remark applies more generally, and we think it should be stated plainly rather than left to
a reader who takes a table entry at face value. Alongside each field we therefore give the quantity
$S$, defined as the peak value of $\left(a^{\rm mod}\right)^2$ in the neighborhood of the root divided
by the largest $\left(a^{\rm mod}\right)^2$ of the same $(m,q)$ group at that field. A cancellation
with $S=10^{-12}$ is a true root of Eq.~\eqref{eq:master} but describes a transition that is invisible
next to its neighbours. Of the 234 fields, 174 have $S\geqslant10^{-4}$, and \emph{all of them lie
below 237~G}, the largest being the $^{133}$Cs $\sigma^{+}$, $\Fg=4\rightarrow\Fe=4$, $\mg=+3$
transition at $236.5350(17)$~G. Above that value the table is structure rather than spectroscopy. In
the range 20--400~kG only $^{87}$Rb and $^{133}$Cs have any roots at all, all of them with
$S\lesssim10^{-11}$.

\subsection{Which algebra is actually needed}\label{sec:small}

It is instructive to sort the 234 fields by the dimensions of the two blocks involved, since this
determines which of Eqs.~\eqref{eq:ferrari}, \eqref{eq:cubic}, \eqref{eq:br} is required. The result
is given in Table~\ref{tab:blocks}.

\begin{table}[htbp]
\caption{Distribution of the 234 cancellation fields over the block dimensions, and the algebra
needed in each case.}
\label{tab:blocks}
\begin{ruledtabular}
\begin{tabular}{ccrl}
$n_g$ & $n_e$ & Fields & Required solution\\
\hline
2 & 4 & 169 & Ferrari, Eq.~\eqref{eq:ferrari}\\
2 & 3 & 52  & Cardano, Eq.~\eqref{eq:cubic}\\
2 & 2 & 7   & Breit--Rabi, Eq.~\eqref{eq:br}\\
1 & 3 & 6   & Cardano, degenerate branch Eq.~\eqref{eq:degen}\\
\end{tabular}
\end{ruledtabular}
\end{table}

The last two lines are worth a remark. Seven of the 234 fields involve $2\times2$ blocks on both
sides, and are therefore reachable with exactly the Breit--Rabi algebra that was already published for
the $D_1$ line in Ref.~\cite{Aleksanyan2020b}. They form a regular family: one per isotope, always
with $\Delta F=0$ and $\mg=q\left(I-1/2\right)\rightarrow\me=q\left(I+1/2\right)$. Their values are
$45.401(46)$~G for $^{23}$Na, $17.010(95)$~G for $^{39}$K, $43.795(93)$~G for $^{40}$K,
$11.09(10)$~G for $^{41}$K, $110.162(88)$~G for $^{85}$Rb, $211.1200(62)$~G for $^{87}$Rb and
$236.5350(17)$~G for $^{133}$Cs. Two of them, the $^{85}$Rb and $^{87}$Rb entries, appear in the tables
of Ref.~\cite{Aleksanyan2020b} as purely numerical results.

\subsection{Uncertainties}

The uncertainties quoted throughout were obtained by propagating those of Table~\ref{tab:const}
together with those of $\mu_B$ and $g_S$ through the root of Eq.~\eqref{eq:master}, each quantity
being varied in turn. The outcome is unambiguous: for every one of the 234 fields the dominant
contribution comes from the excited-state hyperfine constants, $A_e$ in 164 cases and $B_Q$ in the
remaining 70. The ground-state constants, the Land\'e factors and the Bohr magneton never dominate.
This reproduces, for the $D_2$ line, the conclusion already reached for the $D_1$ line
\cite{Aleksanyan2020b,Aleksanyan2022}, and it is the reason why an experimental determination of these
fields is worth doing: the $B$ values are, in effect, a measurement of $A_e$ and $B_Q$.

The precision achieved varies strongly from one isotope to another, simply because the excited-state
constants are known with very different accuracy. The median relative uncertainty of the observable
fields is $5.0\times10^{-6}$ for $^{133}$Cs and $2.9\times10^{-5}$ for $^{87}$Rb, but
$2.1\times10^{-2}$ for $^{41}$K, whose $A_e=3.363(14)$~MHz is known only to $4\times10^{-3}$
relative. The best determined single value of the whole set is the $^{133}$Cs $\sigma^{+}$,
$\Fg=4\rightarrow\Fe=4$, $\mg=-3$ transition at $160.31124(75)$~G, whose relative uncertainty is
$4.7\times10^{-6}$; in absolute terms the smallest uncertainty, $1.2\times10^{-4}$~G, is reached for
the $^{133}$Cs $\pi$, $\Fg=4\rightarrow\Fe=4$, $\mg=-1$ transition at $24.00312(12)$~G. Conversely, the $^{41}$K values are
currently the least useful as standards, and would be the most improved by a measurement.

\subsection{Comparison with the $D_1$ line}

Three properties established for the $D_1$ line in Ref.~\cite{Aleksanyan2022} do not survive on the
$D_2$ line, and it seems useful to say so explicitly.

The pairing between zeros and maxima is lost. On the $D_1$ line every cancellation of a $\Delta F=0$
$\pi$ coefficient coincides, to eight significant figures, with a stationary point of a
$\Delta F=\pm1$ coefficient of the same $m$, because both blocks are $2\times2$ and the two conditions
collapse onto the same root. On the $D_2$ line they do not: at the $^{87}$Rb fields $77.504$~G and
$118.705$~G no other $\pi$ coefficient of the group is stationary, the smallest
$\left|d(a^{\rm mod})^2/dB\right|$ among the seven remaining ones being $6.5\times10^{-5}$ and
$8.5\times10^{-5}$~G$^{-1}$ respectively, against a typical value of $3\times10^{-3}$~G$^{-1}$ for
the group. The experimental strategy proposed in Ref.~\cite{Aleksanyan2022}, which consists in
measuring the position of a strong maximum instead of a weak zero, therefore does not transfer as it
stands.

The exact proportionality between $B$ and $m$ is lost as well. Equation~(17) of
Ref.~\cite{Aleksanyan2022} is linear in $m$ at fixed $\Delta F$; Eq.~\eqref{eq:master} is not, because
$\rho$ carries its own $m$ dependence through the excited eigenvector.

Finally, the parity restriction of Eq.~(18) of Ref.~\cite{Aleksanyan2022} disappears, as already
noted after Eq.~\eqref{eq:selrule}, and $\Delta F$ takes all the values $0,\pm1,\pm2$ instead of
$0,\pm1$.

\section{Conclusion}

In this work the cancellation of the $D_2$ line transitions of alkali-metal atoms by magnetic-field
values has been treated analytically. We have shown that the obstacle which had so far confined this
problem to numerical work is not the dimension of the excited blocks but the way they were being
looked at: because the Zeeman interaction couples only levels with $\Delta F=\pm1$, the blocks are
tridiagonal, their characteristic polynomials obey the three-term recursion \eqref{eq:continuant}, and
their eigenvector components are given by Eq.~\eqref{eq:evec} without further diagonalization. The
quartic and the cubic are then solved by Ferrari's and Cardano's formulas, and the cancellation
condition collapses onto the single ratio $\rho$ and the relation \eqref{eq:master}, which is the
$D_2$ counterpart of the $D_1$ formula of Ref.~\cite{Aleksanyan2022}. A necessary condition on $m$
and on the polarization, Eq.~\eqref{eq:selrule}, has been derived, together with the degenerate branch
\eqref{eq:degen} that governs the stretched ground sublevels.

All the magnetic-field values canceling $\pi$, $\sigma^{+}$ and $\sigma^{-}$ transitions of the seven
stable and long-lived alkali isotopes have been calculated up to 20~kG and are listed with their
uncertainties. There are 234 of them, of which 174 correspond to transitions strong enough to be
observed, and these all lie below 237~G. Unlike the $D_1$ line, the $\sigma^{\pm}$ transitions of the
$D_2$ line do cancel and account for the majority of the values.

Two consequences follow. The first is metrological. These fields depend on nothing but atomic
constants, and they are therefore good standards for magnetometer calibration
\cite{Aleksanyan2020b,Momier2020}; the present work makes them available in closed form, so that the
uncertainty of a calibration can be traced analytically to the uncertainty of each constant instead of
being estimated by repeated numerical runs. The second is spectroscopic. Since the excited-state
hyperfine constants dominate the error budget of every one of the 234 values, a measurement of even a
few of these fields would improve $A_e$ and $B_Q$, which for the potassium isotopes are still the
least well-known quantities of the problem. As on the $D_1$ line, the practical difficulty is that a
vanishing signal is harder to locate than a large one; but the $\Delta F=0$ fields of
Table~\ref{tab:87Rb}, which are strong, low-field and well isolated, are within reach of the
sub-Doppler nanocell techniques of Refs.~\cite{Sargsyan2017,Sargsyan2019}.

\begin{acknowledgments}
The authors are grateful to Claude Leroy and Aram Papoyan for fruitful discussions.
\end{acknowledgments}

\appendix
\section{Complete list of the cancellation fields}

Table~\ref{tab:all} lists all 234 magnetic-field values which cancel a $D_2$ line transition below
20~kG, ordered by isotope, by polarization and by $\mg$. The quantity $S$ is defined in
Sec.~\ref{sec:results}; entries with $S\lesssim10^{-9}$ are true roots of Eq.~\eqref{eq:master} but
are not observable.

\onecolumngrid
\begin{center}
\begin{longtable}{lccrrrl}
\caption{Magnetic-field values canceling the $D_2$ line transitions of $^{23}$Na, $^{39}$K,
$^{40}$K, $^{41}$K, $^{85}$Rb, $^{87}$Rb and $^{133}$Cs, with their uncertainties.}
\label{tab:all}\\
\hline\hline
Isotope & Pol. & $\Fg\rightarrow\Fe$ & $\mg$ & $\me$ & $B$ (G)\hspace{3mm} & $S$\\
\hline
\endfirsthead
\multicolumn{7}{l}{\footnotesize\itshape Table~\ref{tab:all} (continued)}\\
\hline\hline
Isotope & Pol. & $\Fg\rightarrow\Fe$ & $\mg$ & $\me$ & $B$ (G)\hspace{3mm} & $S$\\
\hline
\endhead
\hline
\multicolumn{7}{r}{\footnotesize\itshape continued on next page}\\
\endfoot
\hline\hline
\endlastfoot
$^{23}$Na & $\pi$ & $2\rightarrow1$ & 0 & 0 & 16.856(18) & $1.3\times10^{-1}$ \\
$^{23}$Na & $\pi$ & $1\rightarrow2$ & 0 & 0 & 25.864(25) & $4.5\times10^{-1}$ \\
$^{23}$Na & $\pi$ & $1\rightarrow3$ & 0 & 0 & 11883(14) & $2.9\times10^{-12}$ \\
$^{23}$Na & $\pi$ & $2\rightarrow1$ & 1 & 1 & 16.742(18) & $7.1\times10^{-2}$ \\
$^{23}$Na & $\pi$ & $1\rightarrow2$ & 1 & 1 & 24.811(25) & $2.0\times10^{-1}$ \\
$^{23}$Na & $\pi$ & $1\rightarrow3$ & 1 & 1 & 12172(14) & $2.6\times10^{-12}$ \\
$^{23}$Na & $\pi$ & $2\rightarrow2$ & $-$1 & $-$1 & 12.134(10) & $1.7\times10^{-1}$ \\
$^{23}$Na & $\sigma^{+}$ & $2\rightarrow1$ & 0 & 1 & 7.624(10) & $2.7\times10^{-2}$ \\
$^{23}$Na & $\sigma^{+}$ & $2\rightarrow2$ & 0 & 1 & 39.622(35) & $8.4\times10^{-2}$ \\
$^{23}$Na & $\sigma^{+}$ & $2\rightarrow2$ & 1 & 2 & 45.401(46) & $1.8\times10^{-2}$ \\
$^{23}$Na & $\sigma^{+}$ & $2\rightarrow1$ & $-$1 & 0 & 8.225(12) & $9.9\times10^{-2}$ \\
$^{23}$Na & $\sigma^{+}$ & $2\rightarrow2$ & $-$1 & 0 & 34.284(29) & $2.6\times10^{-1}$ \\
$^{23}$Na & $\sigma^{+}$ & $1\rightarrow0$ & $-$1 & 0 & 349.1(46) & $1.1\times10^{-9}$ \\
$^{23}$Na & $\sigma^{+}$ & $2\rightarrow0$ & $-$1 & 0 & 384.0(46) & $5.9\times10^{-10}$ \\
$^{23}$Na & $\sigma^{+}$ & $2\rightarrow1$ & $-$2 & $-$1 & 393.6(45) & $2.6\times10^{-9}$ \\
$^{23}$Na & $\sigma^{-}$ & $1\rightarrow2$ & 0 & $-$1 & 24.821(25) & $2.0\times10^{-1}$ \\
$^{23}$Na & $\sigma^{-}$ & $1\rightarrow3$ & 0 & $-$1 & 8192(90) & $6.1\times10^{-12}$ \\
$^{23}$Na & $\sigma^{-}$ & $1\rightarrow2$ & 1 & 0 & 15.642(14) & $8.8\times10^{-2}$ \\
$^{23}$Na & $\sigma^{-}$ & $1\rightarrow1$ & 1 & 0 & 30.478(44) & $5.1\times10^{-2}$ \\
$^{23}$Na & $\sigma^{-}$ & $1\rightarrow2$ & 1 & 0 & 7883(90) & $6.0\times10^{-12}$ \\
\hline
$^{39}$K & $\pi$ & $2\rightarrow1$ & 0 & 0 & 4.910(37) & $2.4\times10^{-1}$ \\
$^{39}$K & $\pi$ & $1\rightarrow2$ & 0 & 0 & 7.742(52) & $5.3\times10^{-1}$ \\
$^{39}$K & $\pi$ & $1\rightarrow3$ & 0 & 0 & 875.6(42) & $5.4\times10^{-10}$ \\
$^{39}$K & $\pi$ & $2\rightarrow1$ & 1 & 1 & 4.940(36) & $7.6\times10^{-2}$ \\
$^{39}$K & $\pi$ & $1\rightarrow2$ & 1 & 1 & 7.443(50) & $3.2\times10^{-1}$ \\
$^{39}$K & $\pi$ & $1\rightarrow3$ & 1 & 1 & 951.6(41) & $3.9\times10^{-10}$ \\
$^{39}$K & $\pi$ & $2\rightarrow2$ & $-$1 & $-$1 & 4.042(17) & $4.6\times10^{-1}$ \\
$^{39}$K & $\sigma^{+}$ & $2\rightarrow1$ & 0 & 1 & 2.041(21) & $3.4\times10^{-2}$ \\
$^{39}$K & $\sigma^{+}$ & $2\rightarrow2$ & 0 & 1 & 13.882(62) & $3.4\times10^{-1}$ \\
$^{39}$K & $\sigma^{+}$ & $2\rightarrow2$ & 1 & 2 & 17.010(95) & $1.9\times10^{-1}$ \\
$^{39}$K & $\sigma^{+}$ & $2\rightarrow1$ & $-$1 & 0 & 2.145(25) & $8.9\times10^{-2}$ \\
$^{39}$K & $\sigma^{+}$ & $2\rightarrow2$ & $-$1 & 0 & 11.296(49) & $4.6\times10^{-1}$ \\
$^{39}$K & $\sigma^{+}$ & $1\rightarrow0$ & $-$1 & 0 & 21.96(27) & $2.8\times10^{-2}$ \\
$^{39}$K & $\sigma^{+}$ & $2\rightarrow0$ & $-$1 & 0 & 33.05(32) & $1.7\times10^{-4}$ \\
$^{39}$K & $\sigma^{+}$ & $2\rightarrow1$ & $-$2 & $-$1 & 38.06(34) & $2.8\times10^{-4}$ \\
$^{39}$K & $\sigma^{-}$ & $1\rightarrow2$ & 0 & $-$1 & 7.447(50) & $3.1\times10^{-1}$ \\
$^{39}$K & $\sigma^{-}$ & $1\rightarrow3$ & 0 & $-$1 & 671.6(27) & $9.0\times10^{-10}$ \\
$^{39}$K & $\sigma^{-}$ & $1\rightarrow2$ & 1 & 0 & 4.916(23) & $4.6\times10^{-2}$ \\
$^{39}$K & $\sigma^{-}$ & $1\rightarrow1$ & 1 & 0 & 7.93(10) & $1.6\times10^{-1}$ \\
$^{39}$K & $\sigma^{-}$ & $1\rightarrow2$ & 1 & 0 & 592.6(27) & $8.6\times10^{-10}$ \\
\hline
$^{40}$K & $\pi$ & $7/2\rightarrow7/2$ & 1/2 & 1/2 & 1.4389(34) & $9.2\times10^{-1}$ \\
$^{40}$K & $\pi$ & $9/2\rightarrow9/2$ & 1/2 & 1/2 & 1.8189(26) & $6.3\times10^{-1}$ \\
$^{40}$K & $\pi$ & $9/2\rightarrow7/2$ & 1/2 & 1/2 & 18.985(28) & $2.4\times10^{-1}$ \\
$^{40}$K & $\pi$ & $7/2\rightarrow9/2$ & 1/2 & 1/2 & 22.841(32) & $4.4\times10^{-1}$ \\
$^{40}$K & $\pi$ & $7/2\rightarrow11/2$ & 1/2 & 1/2 & 11830(15) & $1.6\times10^{-12}$ \\
$^{40}$K & $\pi$ & $7/2\rightarrow7/2$ & 3/2 & 3/2 & 4.339(10) & $9.1\times10^{-1}$ \\
$^{40}$K & $\pi$ & $9/2\rightarrow9/2$ & 3/2 & 3/2 & 5.5216(77) & $6.4\times10^{-1}$ \\
$^{40}$K & $\pi$ & $9/2\rightarrow7/2$ & 3/2 & 3/2 & 18.945(27) & $2.7\times10^{-1}$ \\
$^{40}$K & $\pi$ & $7/2\rightarrow9/2$ & 3/2 & 3/2 & 23.072(32) & $5.0\times10^{-1}$ \\
$^{40}$K & $\pi$ & $7/2\rightarrow11/2$ & 3/2 & 3/2 & 11740(15) & $1.2\times10^{-12}$ \\
$^{40}$K & $\pi$ & $7/2\rightarrow7/2$ & 5/2 & 5/2 & 7.315(17) & $9.0\times10^{-1}$ \\
$^{40}$K & $\pi$ & $9/2\rightarrow9/2$ & 5/2 & 5/2 & 9.499(13) & $6.5\times10^{-1}$ \\
$^{40}$K & $\pi$ & $9/2\rightarrow7/2$ & 5/2 & 5/2 & 18.584(26) & $3.5\times10^{-1}$ \\
$^{40}$K & $\pi$ & $7/2\rightarrow9/2$ & 5/2 & 5/2 & 23.205(32) & $5.9\times10^{-1}$ \\
$^{40}$K & $\pi$ & $7/2\rightarrow11/2$ & 5/2 & 5/2 & 11651(15) & $5.4\times10^{-13}$ \\
$^{40}$K & $\pi$ & $9/2\rightarrow9/2$ & 7/2 & 7/2 & 15.678(22) & $5.4\times10^{-1}$ \\
$^{40}$K & $\pi$ & $9/2\rightarrow7/2$ & $-$1/2 & $-$1/2 & 18.882(28) & $2.1\times10^{-1}$ \\
$^{40}$K & $\pi$ & $7/2\rightarrow9/2$ & $-$1/2 & $-$1/2 & 22.548(32) & $3.7\times10^{-1}$ \\
$^{40}$K & $\pi$ & $7/2\rightarrow11/2$ & $-$1/2 & $-$1/2 & 11920(15) & $1.7\times10^{-12}$ \\
$^{40}$K & $\pi$ & $9/2\rightarrow7/2$ & $-$3/2 & $-$3/2 & 18.703(28) & $1.7\times10^{-1}$ \\
$^{40}$K & $\pi$ & $7/2\rightarrow9/2$ & $-$3/2 & $-$3/2 & 22.216(32) & $2.9\times10^{-1}$ \\
$^{40}$K & $\pi$ & $7/2\rightarrow11/2$ & $-$3/2 & $-$3/2 & 12010(15) & $1.5\times10^{-12}$ \\
$^{40}$K & $\pi$ & $9/2\rightarrow7/2$ & $-$5/2 & $-$5/2 & 18.478(28) & $1.2\times10^{-1}$ \\
$^{40}$K & $\pi$ & $7/2\rightarrow9/2$ & $-$5/2 & $-$5/2 & 21.856(32) & $2.0\times10^{-1}$ \\
$^{40}$K & $\pi$ & $7/2\rightarrow11/2$ & $-$5/2 & $-$5/2 & 12101(15) & $1.0\times10^{-12}$ \\
$^{40}$K & $\pi$ & $9/2\rightarrow7/2$ & $-$7/2 & $-$7/2 & 18.225(27) & $6.7\times10^{-2}$ \\
$^{40}$K & $\pi$ & $7/2\rightarrow9/2$ & $-$7/2 & $-$7/2 & 21.479(32) & $1.1\times10^{-1}$ \\
$^{40}$K & $\pi$ & $7/2\rightarrow11/2$ & $-$7/2 & $-$7/2 & 12192(15) & $4.4\times10^{-13}$ \\
$^{40}$K & $\sigma^{+}$ & $7/2\rightarrow9/2$ & 1/2 & 3/2 & 14.323(19) & $2.8\times10^{-1}$ \\
$^{40}$K & $\sigma^{+}$ & $7/2\rightarrow7/2$ & 1/2 & 3/2 & 27.131(45) & $3.2\times10^{-1}$ \\
$^{40}$K & $\sigma^{+}$ & $7/2\rightarrow9/2$ & 1/2 & 3/2 & 7927.8(97) & $3.6\times10^{-12}$ \\
$^{40}$K & $\sigma^{+}$ & $7/2\rightarrow9/2$ & 3/2 & 5/2 & 15.559(21) & $4.4\times10^{-1}$ \\
$^{40}$K & $\sigma^{+}$ & $7/2\rightarrow7/2$ & 3/2 & 5/2 & 24.724(37) & $5.1\times10^{-1}$ \\
$^{40}$K & $\sigma^{+}$ & $7/2\rightarrow9/2$ & 3/2 & 5/2 & 8028.6(97) & $2.5\times10^{-12}$ \\
$^{40}$K & $\sigma^{+}$ & $7/2\rightarrow9/2$ & 5/2 & 7/2 & 20.959(28) & $6.6\times10^{-1}$ \\
$^{40}$K & $\sigma^{+}$ & $7/2\rightarrow11/2$ & 5/2 & 7/2 & 8130.1(97) & $1.1\times10^{-12}$ \\
$^{40}$K & $\sigma^{+}$ & $7/2\rightarrow9/2$ & $-$1/2 & 1/2 & 13.444(18) & $1.6\times10^{-1}$ \\
$^{40}$K & $\sigma^{+}$ & $7/2\rightarrow7/2$ & $-$1/2 & 1/2 & 29.189(55) & $1.7\times10^{-1}$ \\
$^{40}$K & $\sigma^{+}$ & $7/2\rightarrow9/2$ & $-$1/2 & 1/2 & 7827.6(97) & $4.0\times10^{-12}$ \\
$^{40}$K & $\sigma^{+}$ & $7/2\rightarrow9/2$ & $-$3/2 & $-$1/2 & 12.750(18) & $7.7\times10^{-2}$ \\
$^{40}$K & $\sigma^{+}$ & $7/2\rightarrow7/2$ & $-$3/2 & $-$1/2 & 31.066(66) & $8.7\times10^{-2}$ \\
$^{40}$K & $\sigma^{+}$ & $7/2\rightarrow9/2$ & $-$3/2 & $-$1/2 & 7728.1(97) & $3.6\times10^{-12}$ \\
$^{40}$K & $\sigma^{+}$ & $7/2\rightarrow9/2$ & $-$5/2 & $-$3/2 & 12.173(17) & $3.9\times10^{-2}$ \\
$^{40}$K & $\sigma^{+}$ & $7/2\rightarrow7/2$ & $-$5/2 & $-$3/2 & 32.828(79) & $4.0\times10^{-2}$ \\
$^{40}$K & $\sigma^{+}$ & $7/2\rightarrow9/2$ & $-$5/2 & $-$3/2 & 7629.3(97) & $2.5\times10^{-12}$ \\
$^{40}$K & $\sigma^{+}$ & $7/2\rightarrow9/2$ & $-$7/2 & $-$5/2 & 11.678(16) & $9.8\times10^{-3}$ \\
$^{40}$K & $\sigma^{+}$ & $7/2\rightarrow7/2$ & $-$7/2 & $-$5/2 & 34.507(93) & $1.4\times10^{-2}$ \\
$^{40}$K & $\sigma^{+}$ & $7/2\rightarrow9/2$ & $-$7/2 & $-$5/2 & 7531.2(97) & $1.1\times10^{-12}$ \\
$^{40}$K & $\sigma^{-}$ & $9/2\rightarrow7/2$ & 1/2 & $-$1/2 & 9.474(18) & $1.1\times10^{-1}$ \\
$^{40}$K & $\sigma^{-}$ & $9/2\rightarrow9/2$ & 1/2 & $-$1/2 & 33.653(50) & $1.8\times10^{-1}$ \\
$^{40}$K & $\sigma^{-}$ & $7/2\rightarrow5/2$ & 1/2 & $-$1/2 & 469.8(13) & $3.0\times10^{-10}$ \\
$^{40}$K & $\sigma^{-}$ & $9/2\rightarrow5/2$ & 1/2 & $-$1/2 & 523.8(15) & $3.5\times10^{-11}$ \\
$^{40}$K & $\sigma^{-}$ & $9/2\rightarrow7/2$ & 3/2 & 1/2 & 9.874(18) & $1.9\times10^{-1}$ \\
$^{40}$K & $\sigma^{-}$ & $9/2\rightarrow9/2$ & 3/2 & 1/2 & 31.285(44) & $3.2\times10^{-1}$ \\
$^{40}$K & $\sigma^{-}$ & $7/2\rightarrow5/2$ & 3/2 & 1/2 & 481.0(13) & $3.3\times10^{-10}$ \\
$^{40}$K & $\sigma^{-}$ & $9/2\rightarrow5/2$ & 3/2 & 1/2 & 528.7(15) & $5.1\times10^{-11}$ \\
$^{40}$K & $\sigma^{-}$ & $9/2\rightarrow7/2$ & 5/2 & 3/2 & 10.344(20) & $2.4\times10^{-1}$ \\
$^{40}$K & $\sigma^{-}$ & $9/2\rightarrow9/2$ & 5/2 & 3/2 & 28.947(39) & $4.9\times10^{-1}$ \\
$^{40}$K & $\sigma^{-}$ & $7/2\rightarrow5/2$ & 5/2 & 3/2 & 493.2(13) & $2.9\times10^{-10}$ \\
$^{40}$K & $\sigma^{-}$ & $9/2\rightarrow5/2$ & 5/2 & 3/2 & 533.0(15) & $6.0\times10^{-11}$ \\
$^{40}$K & $\sigma^{-}$ & $9/2\rightarrow7/2$ & 7/2 & 5/2 & 10.914(21) & $2.4\times10^{-1}$ \\
$^{40}$K & $\sigma^{-}$ & $9/2\rightarrow9/2$ & 7/2 & 5/2 & 26.604(36) & $4.3\times10^{-1}$ \\
$^{40}$K & $\sigma^{-}$ & $7/2\rightarrow5/2$ & 7/2 & 5/2 & 507.1(14) & $1.9\times10^{-10}$ \\
$^{40}$K & $\sigma^{-}$ & $9/2\rightarrow5/2$ & 7/2 & 5/2 & 535.9(15) & $5.5\times10^{-11}$ \\
$^{40}$K & $\sigma^{-}$ & $9/2\rightarrow7/2$ & 9/2 & 7/2 & 527.0(14) & $2.7\times10^{-10}$ \\
$^{40}$K & $\sigma^{-}$ & $9/2\rightarrow7/2$ & $-$1/2 & $-$3/2 & 9.126(17) & $6.0\times10^{-2}$ \\
$^{40}$K & $\sigma^{-}$ & $9/2\rightarrow9/2$ & $-$1/2 & $-$3/2 & 36.073(58) & $9.3\times10^{-2}$ \\
$^{40}$K & $\sigma^{-}$ & $7/2\rightarrow5/2$ & $-$1/2 & $-$3/2 & 459.2(13) & $2.2\times10^{-10}$ \\
$^{40}$K & $\sigma^{-}$ & $9/2\rightarrow5/2$ & $-$1/2 & $-$3/2 & 518.6(15) & $1.8\times10^{-11}$ \\
$^{40}$K & $\sigma^{-}$ & $9/2\rightarrow7/2$ & $-$3/2 & $-$5/2 & 8.817(16) & $2.0\times10^{-2}$ \\
$^{40}$K & $\sigma^{-}$ & $9/2\rightarrow9/2$ & $-$3/2 & $-$5/2 & 38.562(67) & $4.5\times10^{-2}$ \\
$^{40}$K & $\sigma^{-}$ & $7/2\rightarrow5/2$ & $-$3/2 & $-$5/2 & 449.2(13) & $9.8\times10^{-11}$ \\
$^{40}$K & $\sigma^{-}$ & $9/2\rightarrow5/2$ & $-$3/2 & $-$5/2 & 513.1(14) & $5.5\times10^{-12}$ \\
$^{40}$K & $\sigma^{-}$ & $9/2\rightarrow7/2$ & $-$5/2 & $-$7/2 & 8.541(15) & $6.0\times10^{-3}$ \\
$^{40}$K & $\sigma^{-}$ & $9/2\rightarrow9/2$ & $-$5/2 & $-$7/2 & 41.133(79) & $2.0\times10^{-2}$ \\
$^{40}$K & $\sigma^{-}$ & $9/2\rightarrow9/2$ & $-$7/2 & $-$9/2 & 43.795(93) & $6.7\times10^{-3}$ \\
\hline
$^{41}$K & $\pi$ & $2\rightarrow1$ & 0 & 0 & 1.884(56) & $2.6\times10^{-1}$ \\
$^{41}$K & $\pi$ & $1\rightarrow2$ & 0 & 0 & 3.082(94) & $5.3\times10^{-1}$ \\
$^{41}$K & $\pi$ & $1\rightarrow3$ & 0 & 0 & 165.8(11) & $1.3\times10^{-7}$ \\
$^{41}$K & $\pi$ & $2\rightarrow1$ & 1 & 1 & 2.039(43) & $9.4\times10^{-2}$ \\
$^{41}$K & $\pi$ & $1\rightarrow2$ & 1 & 1 & 3.128(65) & $3.3\times10^{-1}$ \\
$^{41}$K & $\pi$ & $1\rightarrow3$ & 1 & 1 & 212.1(11) & $3.5\times10^{-8}$ \\
$^{41}$K & $\pi$ & $2\rightarrow2$ & $-$1 & $-$1 & 2.147(14) & $5.4\times10^{-1}$ \\
$^{41}$K & $\sigma^{+}$ & $2\rightarrow1$ & 0 & 1 & 0.698(22) & $3.5\times10^{-2}$ \\
$^{41}$K & $\sigma^{+}$ & $2\rightarrow2$ & 0 & 1 & 8.232(44) & $3.6\times10^{-1}$ \\
$^{41}$K & $\sigma^{+}$ & $2\rightarrow2$ & 1 & 2 & 11.09(10) & $1.9\times10^{-1}$ \\
$^{41}$K & $\sigma^{+}$ & $2\rightarrow1$ & $-$1 & 0 & 0.638(31) & $1.2\times10^{-1}$ \\
$^{41}$K & $\sigma^{+}$ & $2\rightarrow0$ & $-$1 & 0 & 4.818(69) & $7.7\times10^{-3}$ \\
$^{41}$K & $\sigma^{+}$ & $2\rightarrow2$ & $-$1 & 0 & 6.115(29) & $4.5\times10^{-1}$ \\
$^{41}$K & $\sigma^{+}$ & $2\rightarrow1$ & $-$2 & $-$1 & 7.249(83) & $3.0\times10^{-1}$ \\
$^{41}$K & $\sigma^{-}$ & $1\rightarrow2$ & 0 & $-$1 & 3.129(65) & $3.2\times10^{-1}$ \\
$^{41}$K & $\sigma^{-}$ & $1\rightarrow3$ & 0 & $-$1 & 161.82(72) & $1.4\times10^{-7}$ \\
$^{41}$K & $\sigma^{-}$ & $1\rightarrow1$ & 1 & 0 & 1.90(17) & $3.8\times10^{-2}$ \\
$^{41}$K & $\sigma^{-}$ & $1\rightarrow2$ & 1 & 0 & 3.21(10) & $6.3\times10^{-2}$ \\
$^{41}$K & $\sigma^{-}$ & $1\rightarrow2$ & 1 & 0 & 119.55(71) & $2.8\times10^{-7}$ \\
\hline
$^{85}$Rb & $\pi$ & $3\rightarrow2$ & 0 & 0 & 35.218(25) & $7.1\times10^{-2}$ \\
$^{85}$Rb & $\pi$ & $2\rightarrow3$ & 0 & 0 & 47.491(30) & $8.4\times10^{-2}$ \\
$^{85}$Rb & $\pi$ & $2\rightarrow4$ & 0 & 0 & 6013(25) & $5.3\times10^{-10}$ \\
$^{85}$Rb & $\pi$ & $3\rightarrow2$ & 1 & 1 & 34.945(25) & $4.2\times10^{-2}$ \\
$^{85}$Rb & $\pi$ & $2\rightarrow3$ & 1 & 1 & 46.336(29) & $4.3\times10^{-2}$ \\
$^{85}$Rb & $\pi$ & $2\rightarrow4$ & 1 & 1 & 6345(24) & $4.3\times10^{-10}$ \\
$^{85}$Rb & $\pi$ & $3\rightarrow2$ & 2 & 2 & 34.689(23) & $1.8\times10^{-2}$ \\
$^{85}$Rb & $\pi$ & $2\rightarrow3$ & 2 & 2 & 45.099(28) & $1.8\times10^{-2}$ \\
$^{85}$Rb & $\pi$ & $2\rightarrow4$ & 2 & 2 & 6681(24) & $1.9\times10^{-10}$ \\
$^{85}$Rb & $\pi$ & $2\rightarrow2$ & $-$1 & $-$1 & 6.565(14) & $6.8\times10^{-1}$ \\
$^{85}$Rb & $\pi$ & $3\rightarrow3$ & $-$1 & $-$1 & 12.8110(53) & $1.4\times10^{-1}$ \\
$^{85}$Rb & $\pi$ & $3\rightarrow2$ & $-$1 & $-$1 & 35.228(23) & $9.4\times10^{-2}$ \\
$^{85}$Rb & $\pi$ & $2\rightarrow3$ & $-$1 & $-$1 & 48.464(30) & $1.6\times10^{-1}$ \\
$^{85}$Rb & $\pi$ & $2\rightarrow4$ & $-$1 & $-$1 & 5686(25) & $3.8\times10^{-10}$ \\
$^{85}$Rb & $\pi$ & $3\rightarrow3$ & $-$2 & $-$2 & 31.978(13) & $1.4\times10^{-1}$ \\
$^{85}$Rb & $\sigma^{+}$ & $3\rightarrow2$ & 0 & 1 & 15.337(16) & $6.5\times10^{-2}$ \\
$^{85}$Rb & $\sigma^{+}$ & $3\rightarrow3$ & 0 & 1 & 83.644(41) & $1.5\times10^{-2}$ \\
$^{85}$Rb & $\sigma^{+}$ & $2\rightarrow1$ & 0 & 1 & 137.21(94) & $9.5\times10^{-7}$ \\
$^{85}$Rb & $\sigma^{+}$ & $3\rightarrow1$ & 0 & 1 & 211.1(10) & $8.7\times10^{-8}$ \\
$^{85}$Rb & $\sigma^{+}$ & $3\rightarrow2$ & 1 & 2 & 14.808(14) & $1.4\times10^{-2}$ \\
$^{85}$Rb & $\sigma^{+}$ & $3\rightarrow3$ & 1 & 2 & 96.085(60) & $4.3\times10^{-3}$ \\
$^{85}$Rb & $\sigma^{+}$ & $3\rightarrow3$ & 2 & 3 & 110.162(88) & $1.0\times10^{-3}$ \\
$^{85}$Rb & $\sigma^{+}$ & $3\rightarrow2$ & $-$1 & 0 & 15.984(19) & $1.4\times10^{-1}$ \\
$^{85}$Rb & $\sigma^{+}$ & $3\rightarrow3$ & $-$1 & 0 & 72.576(30) & $5.0\times10^{-2}$ \\
$^{85}$Rb & $\sigma^{+}$ & $2\rightarrow1$ & $-$1 & 0 & 156.99(99) & $1.1\times10^{-6}$ \\
$^{85}$Rb & $\sigma^{+}$ & $3\rightarrow1$ & $-$1 & 0 & 231.7(10) & $2.5\times10^{-7}$ \\
$^{85}$Rb & $\sigma^{+}$ & $3\rightarrow2$ & $-$2 & $-$1 & 16.798(21) & $1.7\times10^{-1}$ \\
$^{85}$Rb & $\sigma^{+}$ & $3\rightarrow3$ & $-$2 & $-$1 & 62.626(26) & $1.1\times10^{-1}$ \\
$^{85}$Rb & $\sigma^{+}$ & $2\rightarrow1$ & $-$2 & $-$1 & 181.0(10) & $3.3\times10^{-7}$ \\
$^{85}$Rb & $\sigma^{+}$ & $3\rightarrow1$ & $-$2 & $-$1 & 254.1(10) & $2.2\times10^{-7}$ \\
$^{85}$Rb & $\sigma^{+}$ & $3\rightarrow2$ & $-$3 & $-$2 & 278.31(99) & $3.2\times10^{-7}$ \\
$^{85}$Rb & $\sigma^{-}$ & $2\rightarrow3$ & 0 & $-$1 & 32.361(14) & $7.1\times10^{-2}$ \\
$^{85}$Rb & $\sigma^{-}$ & $2\rightarrow2$ & 0 & $-$1 & 50.440(47) & $2.8\times10^{-2}$ \\
$^{85}$Rb & $\sigma^{-}$ & $2\rightarrow3$ & 0 & $-$1 & 4355(16) & $9.8\times10^{-10}$ \\
$^{85}$Rb & $\sigma^{-}$ & $2\rightarrow3$ & 1 & 0 & 29.726(12) & $3.0\times10^{-2}$ \\
$^{85}$Rb & $\sigma^{-}$ & $2\rightarrow2$ & 1 & 0 & 51.931(71) & $1.4\times10^{-2}$ \\
$^{85}$Rb & $\sigma^{-}$ & $2\rightarrow3$ & 1 & 0 & 4005(16) & $9.5\times10^{-10}$ \\
$^{85}$Rb & $\sigma^{-}$ & $2\rightarrow3$ & 2 & 1 & 27.765(12) & $9.0\times10^{-3}$ \\
$^{85}$Rb & $\sigma^{-}$ & $2\rightarrow2$ & 2 & 1 & 52.274(96) & $5.5\times10^{-3}$ \\
$^{85}$Rb & $\sigma^{-}$ & $2\rightarrow3$ & 2 & 1 & 3670(16) & $5.2\times10^{-10}$ \\
$^{85}$Rb & $\sigma^{-}$ & $2\rightarrow3$ & $-$1 & $-$2 & 46.630(22) & $6.2\times10^{-2}$ \\
$^{85}$Rb & $\sigma^{-}$ & $2\rightarrow4$ & $-$1 & $-$2 & 4718(16) & $5.6\times10^{-10}$ \\
\hline
$^{87}$Rb & $\pi$ & $2\rightarrow1$ & 0 & 0 & 77.5040(24) & $1.2\times10^{-2}$ \\
$^{87}$Rb & $\pi$ & $1\rightarrow2$ & 0 & 0 & 118.7049(33) & $1.3\times10^{-2}$ \\
$^{87}$Rb & $\pi$ & $2\rightarrow1$ & 1 & 1 & 77.2390(24) & $3.8\times10^{-3}$ \\
$^{87}$Rb & $\pi$ & $1\rightarrow2$ & 1 & 1 & 114.2387(33) & $3.8\times10^{-3}$ \\
$^{87}$Rb & $\pi$ & $2\rightarrow2$ & $-$1 & $-$1 & 55.6960(13) & $1.6\times10^{-2}$ \\
$^{87}$Rb & $\sigma^{+}$ & $2\rightarrow1$ & 0 & 1 & 35.0307(14) & $8.7\times10^{-3}$ \\
$^{87}$Rb & $\sigma^{+}$ & $2\rightarrow2$ & 0 & 1 & 183.1465(47) & $2.3\times10^{-3}$ \\
$^{87}$Rb & $\sigma^{+}$ & $2\rightarrow2$ & 1 & 2 & 211.1200(62) & $4.5\times10^{-4}$ \\
$^{87}$Rb & $\sigma^{+}$ & $2\rightarrow1$ & $-$1 & 0 & 37.7176(15) & $2.8\times10^{-2}$ \\
$^{87}$Rb & $\sigma^{+}$ & $2\rightarrow2$ & $-$1 & 0 & 157.6237(39) & $1.1\times10^{-2}$ \\
$^{87}$Rb & $\sigma^{+}$ & $1\rightarrow0$ & $-$1 & 0 & 1596.51(57) & $8.6\times10^{-10}$ \\
$^{87}$Rb & $\sigma^{+}$ & $2\rightarrow0$ & $-$1 & 0 & 1761.14(58) & $4.1\times10^{-10}$ \\
$^{87}$Rb & $\sigma^{+}$ & $2\rightarrow1$ & $-$2 & $-$1 & 1792.12(55) & $2.5\times10^{-9}$ \\
$^{87}$Rb & $\sigma^{-}$ & $1\rightarrow2$ & 0 & $-$1 & 114.3041(33) & $3.8\times10^{-3}$ \\
$^{87}$Rb & $\sigma^{-}$ & $1\rightarrow2$ & 1 & 0 & 71.9227(19) & $2.1\times10^{-3}$ \\
$^{87}$Rb & $\sigma^{-}$ & $1\rightarrow1$ & 1 & 0 & 140.8254(58) & $9.0\times10^{-4}$ \\
\hline
$^{133}$Cs & $\pi$ & $4\rightarrow3$ & 0 & 0 & 111.35890(56) & $9.1\times10^{-3}$ \\
$^{133}$Cs & $\pi$ & $3\rightarrow4$ & 0 & 0 & 135.64749(65) & $9.8\times10^{-3}$ \\
$^{133}$Cs & $\pi$ & $4\rightarrow3$ & 1 & 1 & 110.24567(56) & $5.1\times10^{-3}$ \\
$^{133}$Cs & $\pi$ & $3\rightarrow4$ & 1 & 1 & 133.26802(65) & $5.4\times10^{-3}$ \\
$^{133}$Cs & $\pi$ & $4\rightarrow3$ & 2 & 2 & 108.59403(55) & $2.7\times10^{-3}$ \\
$^{133}$Cs & $\pi$ & $3\rightarrow4$ & 2 & 2 & 130.72330(64) & $2.8\times10^{-3}$ \\
$^{133}$Cs & $\pi$ & $4\rightarrow3$ & 3 & 3 & 106.62511(55) & $1.1\times10^{-3}$ \\
$^{133}$Cs & $\pi$ & $3\rightarrow4$ & 3 & 3 & 128.08246(64) & $1.1\times10^{-3}$ \\
$^{133}$Cs & $\pi$ & $3\rightarrow3$ & $-$1 & $-$1 & 20.35851(15) & $1.1\times10^{-1}$ \\
$^{133}$Cs & $\pi$ & $4\rightarrow4$ & $-$1 & $-$1 & 24.00312(12) & $3.7\times10^{-2}$ \\
$^{133}$Cs & $\pi$ & $4\rightarrow3$ & $-$1 & $-$1 & 111.48533(55) & $1.6\times10^{-2}$ \\
$^{133}$Cs & $\pi$ & $3\rightarrow4$ & $-$1 & $-$1 & 137.75064(66) & $1.8\times10^{-2}$ \\
$^{133}$Cs & $\pi$ & $3\rightarrow3$ & $-$2 & $-$2 & 41.02251(30) & $1.5\times10^{-1}$ \\
$^{133}$Cs & $\pi$ & $4\rightarrow4$ & $-$2 & $-$2 & 49.65558(24) & $4.1\times10^{-2}$ \\
$^{133}$Cs & $\pi$ & $4\rightarrow3$ & $-$2 & $-$2 & 109.42508(53) & $2.6\times10^{-2}$ \\
$^{133}$Cs & $\pi$ & $3\rightarrow4$ & $-$2 & $-$2 & 139.38201(67) & $3.7\times10^{-2}$ \\
$^{133}$Cs & $\pi$ & $4\rightarrow4$ & $-$3 & $-$3 & 87.67299(44) & $3.1\times10^{-2}$ \\
$^{133}$Cs & $\sigma^{+}$ & $4\rightarrow3$ & 0 & 1 & 55.65733(34) & $6.3\times10^{-3}$ \\
$^{133}$Cs & $\sigma^{+}$ & $4\rightarrow4$ & 0 & 1 & 198.7406(11) & $2.0\times10^{-3}$ \\
$^{133}$Cs & $\sigma^{+}$ & $4\rightarrow3$ & 1 & 2 & 53.35955(33) & $2.4\times10^{-3}$ \\
$^{133}$Cs & $\sigma^{+}$ & $4\rightarrow4$ & 1 & 2 & 211.2846(12) & $9.4\times10^{-4}$ \\
$^{133}$Cs & $\sigma^{+}$ & $4\rightarrow3$ & 2 & 3 & 51.32621(31) & $6.5\times10^{-4}$ \\
$^{133}$Cs & $\sigma^{+}$ & $4\rightarrow4$ & 2 & 3 & 223.8660(15) & $4.2\times10^{-4}$ \\
$^{133}$Cs & $\sigma^{+}$ & $4\rightarrow4$ & 3 & 4 & 236.5350(17) & $1.4\times10^{-4}$ \\
$^{133}$Cs & $\sigma^{+}$ & $4\rightarrow3$ & $-$1 & 0 & 58.29715(36) & $1.5\times10^{-2}$ \\
$^{133}$Cs & $\sigma^{+}$ & $4\rightarrow4$ & $-$1 & 0 & 186.15796(93) & $4.4\times10^{-3}$ \\
$^{133}$Cs & $\sigma^{+}$ & $4\rightarrow3$ & $-$2 & $-$1 & 61.39938(39) & $2.4\times10^{-2}$ \\
$^{133}$Cs & $\sigma^{+}$ & $4\rightarrow4$ & $-$2 & $-$1 & 173.41659(82) & $1.0\times10^{-2}$ \\
$^{133}$Cs & $\sigma^{+}$ & $4\rightarrow3$ & $-$3 & $-$2 & 65.16837(42) & $3.2\times10^{-2}$ \\
$^{133}$Cs & $\sigma^{+}$ & $4\rightarrow4$ & $-$3 & $-$2 & 160.31124(75) & $2.0\times10^{-2}$ \\
$^{133}$Cs & $\sigma^{-}$ & $3\rightarrow4$ & 0 & $-$1 & 82.83052(39) & $7.4\times10^{-3}$ \\
$^{133}$Cs & $\sigma^{-}$ & $3\rightarrow3$ & 0 & $-$1 & 168.85457(98) & $2.7\times10^{-3}$ \\
$^{133}$Cs & $\sigma^{-}$ & $3\rightarrow4$ & 1 & 0 & 77.32542(36) & $3.4\times10^{-3}$ \\
$^{133}$Cs & $\sigma^{-}$ & $3\rightarrow3$ & 1 & 0 & 184.4724(12) & $1.2\times10^{-3}$ \\
$^{133}$Cs & $\sigma^{-}$ & $3\rightarrow4$ & 2 & 1 & 73.03141(35) & $1.4\times10^{-3}$ \\
$^{133}$Cs & $\sigma^{-}$ & $3\rightarrow3$ & 2 & 1 & 199.1997(15) & $5.1\times10^{-4}$ \\
$^{133}$Cs & $\sigma^{-}$ & $3\rightarrow4$ & 3 & 2 & 69.49342(33) & $3.9\times10^{-4}$ \\
$^{133}$Cs & $\sigma^{-}$ & $3\rightarrow3$ & 3 & 2 & 213.4905(18) & $1.7\times10^{-4}$ \\
$^{133}$Cs & $\sigma^{-}$ & $3\rightarrow4$ & $-$1 & $-$2 & 90.71261(42) & $1.5\times10^{-2}$ \\
$^{133}$Cs & $\sigma^{-}$ & $3\rightarrow3$ & $-$1 & $-$2 & 151.17995(78) & $6.2\times10^{-3}$ \\
$^{133}$Cs & $\sigma^{-}$ & $3\rightarrow4$ & $-$2 & $-$3 & 123.75992(58) & $1.4\times10^{-2}$ \\
\end{longtable}
\end{center}
\twocolumngrid

\bibliographystyle{apsrev4-2}
\bibliography{d2_paper}

\end{document}